# Domain structure in the monoclinic *Pm* phase of Pb(Mg$_{1/3}$Nb$_{2/3}$)O$_3$ – PbTiO$_3$ single crystals

A. A. Bokov and Z.-G. Ye

*Department of Chemistry, Simon Fraser University, Burnaby, BC, V5A 1S6, Canada*

The domain structure of (1-$x$)Pb(Mg$_{1/3}$Nb$_{2/3}$)O$_3$- $x$PbTiO$_3$ single crystals with composition $x \approx 0.33$ in the range of the morphotropic phase boundary (MPB) was studied. Based on the analysis of spontaneous strain compatibility and charge of domain walls, we have established the permissible domain arrangements for the ferroelectric phases of different symmetry, which are expected to occur in the range of the MPB. Examination of (001)-oriented unpoled and electrically poled (along the [001] direction) crystal plates in polarizing microscope reveals a monophase state with the domain structure compatible with the structure theoretically predicted for the $M_C$ monoclinic phase (space group *Pm*), which was recently discovered in the compositions close to the MPB by x-ray and neutron diffraction studies. In the unpoled crystal the 180° walls between the domains whose spontaneous polarization vectors are parallel to the plane of the crystal plate (i.e. *a*-domains) are observed. The domain structure of the poled crystal is predominantly composed of crystallographically prominent $W_f$ walls parallel to (001) (i.e. the plane of the crystal plate) and inclined *S* walls parallel to [110] or [1$\bar{1}$0] direction. In poled and unpoled samples the optical rotatory polarization effect is found, which is related to the inhomogeneity of the sample resulting from the mixture of domains.

PACS numbers: 77.80.Dj, 77.84.Dy, 78.20.Ek

## I. INTRODUCTION

The properties of relaxor ferroelectrics-based solid solutions with perovskite-type structure, (1-$x$)Pb(Mg$_{1/3}$Nb$_{2/3}$)O$_3$- $x$PbTiO$_3$ (PMN-PT), and related materials have become a topic of intensive fundamental and technological interest during the last few years because of their extraordinary piezoelectric performance.[1] It is expected that single crystals of these solid solutions will become the piezoelectric materials of the next generation. Extreme properties are usually related to the closeness of a phase boundary. In the case of piezocrystals, it corresponds to the morphotropic phase boundary (MPB) separating different phases when the composition of the solid solution $x$ varies. In PMN-PT the MPB was located at $x \sim 0.3$ and until recently it had been believed to be the boundary between the rhombohedral *R*3*m* and the tetragonal *P*4*mm* phases. But the latest studies revealed a new monoclinic structure that exists in between the rhombohedral and tetragonal phases.[2-6] Intermediate phases with a monoclinic distortion were also discovered recently near MPB in the other high-piezoelectric perovskite material, (1-$x$)Pb(Zn$_{1/3}$Nb$_{2/3}$)O$_3$ – $x$PbTiO$_3$ (PZN-PT),[5-7] as well as in (1-$x$)PbZrTiO$_3$ – $x$PbTiO$_3$ (PZT),[8] which has been the basic piezoelectric material for many years. Thus, the presence of a monoclinic phase seems to be a common feature for the perovskite solid solutions exhibiting enhanced piezoelectric properties.

Interestingly, monoclinic phases of three different types were found by neutron and x-ray diffraction in PMN- PT. In unpoled samples the phases of space group *Cm* ($M_B$–type)[2] and *Pm* ($M_C$–type)[3,5,6] were observed in the composition ranges of $0.27 < x < 0.3$ and $0.3 < x < 0.35$, respectively (at room temperature). The other monoclinic phase, $M_A$, with the same space group symmetry *Cm*, was found in single crystals poled under a high electric field (35 kV/cm) applied along the pseudocubic [001] direction.[4]

Experiments showed that the best piezoelectric characteristics could be obtained in crystals poled along the <001> direction.[1] But the field applied in this direction is unable to transform the monoclinic or rhombohedral sample into a monodomain state. The crystal should necessarily contain domains of different orientations. Since the dielectric, piezoelectric and many other properties depend closely on domain structure, domain analysis by polarized light microscopy constitutes an important step in the characterization of high-performance piezoelectric materials.

Optical studies of the domains in PMN-PT crystals were reported by several groups, but for the rhombohedral and tetragonal phases only.[9-12] Investigation of the monoclinic phase is a more difficult task because the domains in this phase should be optically biaxial and many more different orientation states are allowed. Domain structure in the crystals containing the monoclinic $Cm$ phase was reported.[13] As for the $Pm$ phase, the first (preliminary) domain patterns of a poled PMN-PT crystal were presented in our recent paper, which was mainly devoted to the dielectric and piezoelectric properties of that phase.[14] In the present work we have studied the monoclinic domain structures in more details with the help of polarized light microscopy both in the poled and virgin states. To interpret the observed domain structure, we have also analyzed the theoretically possible domain configurations for different phases expected in the MPB composition range and concluded that the observed domain wall positions agree well with the predictions made for the $Pm$ phase.

## II. EXPERIMENTAL PROCEDURE AND CRYSTAL CHARACTERIZATION

Single crystals of $(1-x)Pb(Mg_{1/3}Nb_{2/3})O_3 - xPbTiO_3$ with nominal composition $x = 0.35$ grown by the Bridgman method were oriented with the help of a Laue camera. Thin (~ 100 μm) crystal plates were cut so that their large faces (with the dimensions of ~ 2 mm × 3 mm) are perpendicular to the pseudocubic [001] direction (all indexes in this paper are hereafter referred to the cubic system). The crystals were mirror polished and annealed at 600 °C for half an hour to eliminate the stresses induced by polishing. The samples were studied using an Olympus BX60 polarizing microscope. For *in situ* observation of domain structure under dc bias, semitransparent gold layers were sputtered on the large faces of the plate as electrodes. To connect the sample to a high voltage power supply, gold wires were attached to the electrodes by silver paste. The sample was positioned on the microscope stage so that the [001] direction of the crystal, the direction of the polarized light propagation and the direction of the applied electric field were parallel to each other. The experimental set-up is shown schematically in Fig. 1. The poling of a sample was performed by an electric field of 10 kV/cm applied at room temperature or upon cooling through the Curie point. All experiments in this work were performed at room temperature unless otherwise stated.

Crystals of perovskite solid solutions in general and PMN-PT in particular are known to contain macroscopic spatial inhomogeneities of composition $x$, as a result of phase segregation that occurred during the process of crystal growth.[15,16] The real composition can differ significantly from the nominal one. When $x$ is close to the MPB, the composition inhomogeneities could lead to the appearance of different phases in different areas of a crystal. The $T$-$x$ phase diagram of PMN-PT suggests[6] that three ferroelectric phases, namely tetragonal, rhombohedral and monoclinic, are expected to occur in the crystals with the composition close to the MPB. According to symmetry arguments, birefringent domains of the tetragonal and rhombohedral structure should exhibit extinctions along <100> and <110> directions, respectively, when observed in crossed polarizers along [001], but in the monoclinic phase the possible extinction positions are not restricted by the symmetry. In preliminary examinations we observed the crystals with significantly different domain structure, but as we are interested in the recently discovered monoclinic phase only, the samples containing domains with <100> and <110> extinctions were discarded. We also excluded the samples containing significant areas that remain bright at any position of crossed polarizers (these areas may result from the overlapping of the domains of different phases). Before poling, all the areas of the crystals selected for investigation showed clear extinctions in the directions different from either <100> or <110>.

Upon heating, the crystals undergo a transition into another phase that should be the tetragonal one according to the phase diagram obtained by x-ray diffraction.[6] The transition temperature transition varies in the range of 82 - 88 °C for different parts of the sample. In polarizing microscope this high-temperature phase exhibits complex domain structure with extinctions along <100> directions, consistent with the behavior expected for *a*-domains of the tetragonal phase. After poling, the crystal plate remains in complete extinction at any positions between crossed polarizers. It can only be so when the crystal is in the monodomain tetragonal state with the spontaneous polarization vector (and hence the tetragonal axis) perpendicular to the plane of the plate. Therefore, this phase is proved to be indeed tetragonal. At temperatures above 150 – 155 °C an optically isotropic (cubic) phase is observed. The small variation of the phase transition temperature across the crystal suggests some slight composition inhomogeneity which was expected for single crystal samples. The average composition *x* is estimated to be approximately 0.33 by comparison of the observed Curie temperature with the known phase diagram.

## III. DOMAIN STRUCTURE BEFORE POLING

Figure 2 shows a (001) crystal platelet (144 μm thick) observed under the polarizing microscope before the deposition of electrodes. The most part of the plate consists of areas of two kinds [denoted by A and B in Fig. 2 (a)], each of which is characterized by well-defined but different extinction position (see below). A comparatively small part C (located at one of the edges of the crystal plate) exhibits another extinction position. After several heating/cooling cycles through the Curie temperature, the configuration of the A and B areas remained almost unchanged, whereas the boundary between A and C shifted such that C area became significantly smaller. After poling the crystal and subsequently depoling it by heating above the Curie temperature, the configurations of A and B parts were significantly changed while C area completely disappeared. This indicates that C is most probably the same monoclinic phase with different arrangements of domains. This conclusion is also confirmed by the fact that in the poled crystal the domain patterns of all three areas, A, B and C, become undistinguishable (see Sec. IV for more details).

It is found that the extinction of areas A and B can be achieved not only when the polarizer and analyzer are precisely crossed (with the angle $f$ in Fig. 1 equal to 90°), but also when they are slightly decrossed. The extinction positions with crossed configuration are shown in Figs. 2 (a) and 2(b). The extinction occurs in the areas A or B, when the transmission direction of polarizers forms an angle of about +3° or – 3° with the <110> crystallographic direction, respectively. In both cases the extinction is incomplete. When the transmission direction of polarizers is parallel to <110> ($d = 45°$), A and B regions show approximately the same brightness. At this angle $d$, offsetting the analyzer about 3-4° clockwise with respect to the crossed position (i.e. $f \approx 87°$) leads to the extinction of B areas [Fig. 2 (c)]. Offsetting the analyzer about 3-4° in the opposite direction (i.e. $f \approx 93°$) leads to the structure with reversed contrast [i.e. A area is at extinction, but B is not, Fig. 2 (d)]. The extinctions caused by decrossing the polarizers are also incomplete and can be observed at angle $f$ slightly varying across the area A or B.

These observations suggest that the plane of vibration of the light is rotated while transmitting through the crystal. The first possibility to be considered in explaining this fact is the optical activity of the monoclinic *m* phase in PMN-PT. In contrast to the other phases (tetragonal 4*mm* and rhombohedral 3*m*) around the MPB, the crystal symmetry of the *m* phase allows the optical activity.[17] Furthermore, the optical activity was indeed observed in materials with a similar perovskite-type structure: a large effect was reported for the rhombohedral (point group 3) phase of $Pb_{0.92}La_{0.08}(Zr_{0.7}Ti_{0.3})_{0.98}O_3$ transparent ceramics.[18] However, we have obtained some results which are contradictory with the optical activity of the *m* phase. In particular, the rotatory power of such an effect should depend on the wavelength of light,[19] but we have not observed any significant dispersion of extinction angles. For this and for some other reasons discussed below, we suggest that the origin of the rotatory polarization of light discovered in PMN-PT is related to the fine domain substructure of the *m* phase.

Inside A and B areas narrow fibrous domains are visible with curved and unevenly spaced walls running approximately at the same angle $\varphi$ with respect to the [100] direction. This angle $\varphi$ has a value of $24 - 27°$ (it cannot be determined exactly due to the irregularity of the walls) but different signs for A and B domains, so that the [100] axis bisects the angle formed by the domain walls of adjacent A and B areas, as shown in Fig. 3.

When viewed at the diagonal position of crossed polarizers, all the domains of A and B areas display the same forth-order green interference color. The use of a $\lambda/4$ plate or a $\lambda$ plate does not lead to the appearance of any color contrast between the domains belonging to the same area (A or B), but the colors of A and B areas become different. This means that all domains inside the same area (say, A area) are characterized by the same slow (and fast) vibration directions, but between different areas, these directions are different. Optical retardation measured with a Berek compensator gives a value of 1810 nm (in agreement with the interference color), which corresponds to a birefringence value of 0.0126 ($\lambda = 546$ nm).

Note that the domain structure described here is completely different from that reported in Ref. 13 for the (001) cuts of unpoled 0.67PMN-0.33PT crystals which were postulated to be the monoclinic *Cm* phase. The authors of Ref. 13 observed broad straight domains separated by the walls parallel to <100> with the extinction positions either at a variable angle of $5 - 35°$ to <100> (for some domains) or along <110> (for some other domains).

## IV. DOMAIN STRUCTURE IN POLED CRYSTAL

To study the effect of dc electric field on the domain structure, semitransparent gold electrodes were deposited on the large surfaces of the plate and the crystal was poled. Subsequent *in situ* observations did not reveal any noticeable changes in the positions of the domain walls under the applied electric field, which means that the samples we studied had been completely poled. This is in agreement with the fact that the poling field of 10 kV/cm is much higher than the coercive field (see Ref. 14 for more details). The photographs of the domains of the poled crystal observed under polarizing microscope with different magnifications are shown in Fig. 4. The structure is typically composed of domain blocks. Inside each block, birefringent domains are separated by straight stripes, which look darker than the domains, and are directed along [110] (in some blocks) or along [1$\bar{1}$0] (in some others). The width of the birefringent domains is typically in the range of 1-10 μm, but wider domains are observed sometimes. The widths of the interdomain stripes are of the same order of magnitude. As can be clearly seen from Fig. 4 (b), the domains belonging to the different blocks join together along a direction approximately parallel to <100>.

Different brightnesses of domains in crossed polarizers usually indicate different orientations of the optical indicatrix in these domains, but this is not the case for our crystals. When a birefringent (e.g. quarter-wave) plate is superimposed and the crystal is rotated with respect to polarizers, the dark interdomain stripes never become brighter than the domains themselves and complete extinction is observed in the neighboring domains and the stripes between them simultaneously. These observations indicate that the dark stripes are not individual birefringent domains but opaque regions. This is also confirmed by the fact that they are clearly visible not only in crossed polarizers but also without analyzer (see Figs. 5 and 6).

To probe the domain boundaries along the thickness ([001]) direction, high-power objective is focused on different levels of the crystal by moving microscope stage with the crystal plate up or down. During such refocusing, the opaque stripes usually "shift" sideward while retaining their direction along <110>, and their apparent width often changes. This means that their boundaries are inclined and have a complex structure. As an example, the domain patterns at three different levels of the crystal, one located underneath the other, are shown in Fig. 5. On the other hand, regions of the crystal plate can be found where domains and opaque boundaries between them are practically vertical, i.e. parallel to the [001] direction. An example of such a region is shown in Fig. 6 (which was made with a higher magnification).

We also observe in some regions the horizontal layers containing mutually perpendicular opaque stripes, i.e. the stripes along [110] are superimposed on top of the those along [1$\bar{1}$0] direction.

The poled crystal plate consists of different areas of irregular form with dimensions of about 100 μm or larger, each of which is characterized by its own extinction position. In some areas the extinction is incomplete. For the regions with vertical (opaque) interdomain boundaries, the extinction was observed with the vibration directions of crossed polarizers along <100>. For the regions with inclined boundaries, the extinction was observed at some other positions of crossed or decrossed polarizers, without any apparent regularity. An example is shown in Fig. 4 (*a*), which has been made with arbitrarily chosen values of angles *f* and *d*. Some areas in extinction (black areas) are clearly seen, while the other parts are differently colored, indicating spatial variations of birefringence. Presumably, these variations arise from the fact that the light passes through the inclined domain walls of different configurations.

The interference colors and extinction directions in poled crystal are found to depend on external electric field *E*. The typical variation of the extinction direction measured *in situ* as a function of a field applied along the same direction as the poling field is shown in Fig. 7. The dependence of the extinction angle *d* on *E* is linear at low field and tends to plateau at high fields.

Fig. 8 shows the joint of two intersecting domain blocks with different directions of the opaque stripes. Both blocks are in extinction position, but the joint between them is clearly revealed as bright area arising from the birefringence induced by the presence of internal elastic (strain) fields (see Sec. V for the discussion). This is a typical situation observed at the joints of different domain blocks.

## V. ANALYSIS OF POSSIBLE DOMAIN STRUCTURES IN THE MONOCLINIC PHASE AND INTERPRETATION OF THE OBSERVED DOMAIN PATTERNS

Three types of monoclinic *m* phase, $M_A$, $M_B$ and $M_C$, were reported in PMN-PT.[2-6] They can be produced from the same prototypic cubic *m3m* phase by means of slightly different distortions and, consequently, belong to different ferroelectric/ferroelastic species. The monoclinic *Pm* ($M_C$) phase [*m3mFm(p)* species, according to the Aizu notations[20]] can be derived by stretching the original cubic perovskite cell along one of the face diagonals (<110> directions) and by subsequently varying the cubic lattice constants in three crystallographic directions to obtain the three different monoclinic lattice constants $a_m$, $b_m$ and $c_m$. During this transformation, all the symmetry elements of the prototype disappear, except the mirror plane *m* parallel to the {100} plane. In the monoclinic *Cm* ($M_A$ or $M_B$) phase [*m3mFm(s)* species], the other mirror plane parallel to {110} remains after transformation. In this case the distortion consists in the stretch of cubic cell along one of the body diagonals (<111> directions) plus elongation (for $M_A$ phase) or compressions (for $M_B$ phase) of the cubic edge lying in the remaining mirror plane and equal compressions (for $M_A$ phase) or elongations (for $M_B$ phase) of the two other edges.

The analysis of the domain structure expounded below reveals that the symmetry of the PNM-PT crystals studied corresponds to the *Pm*, but not the *Cm*, phase. The possible variants of the prototype cell deformation in the monoclinic *Pm* phase are shown schematically in Fig. 9. Each of the twelve directions shown represents a ferroelastic orientation state and can be considered as parallel to the axis of the optic indicatrix or to the spontaneous polarization ($P_s$) vector (in the monoclinic phase that axis and the $P_s$ vector must lie in the symmetry plane *m*, but their directions must not coincide). For every ferroelastic orientation state the spontaneous polarization can adopt two opposite directions so that the total number of possible orientations for $P_s$ equals to 24. As the number of the possible ferroelectric orientation states (24) is larger than the number of the ferroelastic orientation states (12), the PMN-PT crystals can be classified as *fully ferroelectric and partially ferroelastic*.[20] For the domains of orientation states 9-12 the polarization vector $P_s$ lies within the plane of the

(001) crystal plate, and by analogy with tetragonal crystals, we call them *a*-domains.

Any domain in the crystal should belong to one of the 24 possible orientation states. In a (001) platelet of the *Pm* phase, all domains are viewed under microscope either along an *m* plane or perpendicular to an *m* plane. In both cases the symmetry forbids optical activity in the bulk of the domains.[21] Thus, the rotatory polarization observed in our experiments should be explained in some other way. It is sensible to relate this phenomenon to the presence of domain walls. It was shown previously, when studying the ferroelectric monoclinic phase of try-glycine sulfate, that the mixture of domains and domain walls can lead to an additional contribution to a gyrotropy which is different from the intrinsic optical rotation or electrogyration.[22]

The adjacent domains in ferroelectrics are usually separated by the walls of definite orientations, which satisfy the conditions of mechanical compatibility without additional stresses and electrical neutrality.[23] In real crystals the walls may be misoriented relative to the permissible directions predetermined by mechanical conditions, but this misorientation is usually small. As for electrical conditions, they may not be very strict in the materials with a high electrical conductivity, in which the field of wall charges can be effectively screened.

The ferroelectric domain walls of three types can generally satisfy the mechanical compatibility conditions: (i) crystallographically prominent $W_f$ walls, whose orientation is fixed with respect to the symmetry elements of the prototypic phase, (ii) *S* walls with indices depending on the components of the spontaneous strain tensor, and (iii) arbitrary $W_\infty$ walls between antiparallel domains, any orientation of which is compatible with spontaneous deformation.[23,24] As the spontaneous strain tensor is invariant with respect to $P_s$ reversal, only ferroelastic orientation states need to be considered to find out the directions of $W_f$ and *S* walls satisfying the mechanical compatibility conditions. The charge on the wall is defined by the directions of $P_s$ in adjacent domains (conjugate vectors), so the full number of ferroelectric orientation states should be taken into account to determine the charge. In particular, to be uncharged, $W_\infty$ walls should be parallel to $P_s$ and the conjugate $P_s$ vectors should have a "head-to-tail" arrangement on the $W_f$ and *S* walls.

After poling by an external electric field, the number of orientation states in the crystal decreases, which simplifies of the domain structure to be studied. For this reason we discuss the domain configurations in poled crystal first.

### A. Poled crystal

As mentioned in Introduction, diffraction studies of PMN-PT with MPB compositions revealed a monoclinic $M_A$ phase in crystals poled along [001]. It seems that a stable $M_C$ (or $M_B$) phase could be irreversibly transformed into a metastable $M_A$ phase by a strong enough electric field applied along [001]. Thus, depending on conditions (that are still poorly known, such as poling field strengths and crystal composition), we can expect the existence of both *Pm* and *Cm* phases in the poled crystals. That is why we analyze the domain structure for both cases and then compare the results with the microscopic observations.

If the ferroelectric crystal is poled completely, the number of possible orientations for $P_s$ decreases in comparison with the unpoled case. For the poling field direction used in this work, it is possible for the domains of the *Pm* phase to be polarized in four different directions, denoted by 1, 2, 3 and 4 in Fig. 10 (a). Each of these directions forms the same angle with respect to the poling field and thus can be expected to occur with an equal probability. The numbers of ferroelastic and ferroelectric orientation states are the same here, in contrast to the unpoled crystals where every ferroelastic orientation state can accommodate two oppositely directed polarization vectors.

All the directions of the domain walls, permissible by the mechanical compatibility arguments in ferroic crystals of different symmetry were calculated by Spariel,[25] but only the equations of planes were reported and the correspondence between the equations and the directions of the orientation states in adjacent domains were not specified. Using the same method,[24] we have performed the calculations to find out such a correspondence.[25] The results are presented in Table I, which gives the description of all permissible domain walls in the crystal poled in the [001] (= $z$) direction. The walls are labeled in accordance with the directions of adjacent domains (e.g. 1/2 stands for the wall between domains 1 and 2) and the orientations of the walls are determined in the rectangular coordinates of the prototypic (cubic) phase. For the $Pm$ phase [$m3mFm(p)$ species], five $W_f$ planes of $x = \pm y$, $x=0$, $y=0$ and $z=0$ type and four $S$ planes are found. $S$ walls are described by the equations

$$(a-b)(x \pm y) = \pm 2dz, \quad (1)$$

where $a$, $b$ and $d$ are the components of the spontaneous strain tensor written for the orientation state in which the monoclinic axis is parallel to $x$ [20,24] (i.e. State 1 in Fig. 10),

$$\begin{pmatrix} a & 0 & 0 \\ & b & d \\ & & c \end{pmatrix}, \quad (2)$$

and $a + b + c = 0$ (to make volume unchanged compared to the prototype). The existence of charged walls is energetically unflavored in PMNT because of its low electrical conductivity. The simplest twinning configurations that can be constructed with the uncharged permissible walls are shown in Figs. 10 (b) and (c). Equations of $S$ walls (1) represent the set of planes parallel to [110] or [1$\bar{1}$0] direction. The traces of these walls on $x = 0$ and $y = 0$ planes form an angle of $q_1$ with [010] (=$y$) and [100] (=$x$) directions, respectively. This angle depends on the values of the monoclinic lattice parameters $a_m$, $b_m$ and $\boldsymbol{b}$ (through the spontaneous strain tensor components), so that when the parameters vary, the walls tilt around the <110> directions to which they are parallel. Fig. 10 shows the case where $a_m > b_m$, and as a result, $q_1 < 90°$. In case of $a_m = b_m$, the walls are parallel to the (001) plane ($z = 0$ plane), i.e. $q_1 = 0$. If $a_m < b_m$, the walls are inclined contrariwise (i.e. the angle $q_1$ in Fig. 10 is larger than 90°). When $\boldsymbol{b} \to 90°$, the angle $q_1$ approaches 90° and it becomes exactly equal to 90° in the limiting case of $\boldsymbol{b} = 90°$, which corresponds to another (orthorhombic) phase which has never been observed in perovskite-type ferroelectrics so far. All $S$ walls in poled crystal remain uncharged at any values of monoclinic lattice parameters. Indeed, the bound charge on the wall equals to zero if the components of conjugate $\boldsymbol{P}_s$ vectors along the normal to the wall are the same. The normals to $S$ walls always lie in one of the {110} planes, thus making the same angle with conjugate $\boldsymbol{P}_s$ vectors. As a result, the mentioned components should be the same.

One can see that the 1/2, 2/3, 3/4 and 1/4 walls can be derived one from the other by means of rotation about the four-fold axis parallel to [001] (i.e. by the symmetry operation lost at the phase transition from the prototype to the $Pm$ phase). The same can be said concerning the 1/3 and 2/4 walls.

By the similar way we determined (see Table I) the domain wall positions for the [001]-poled $M_A$ and $M_B$ phases, which have the same $Cm$ symmetry [$m3mFm(s)$ species]. Uncharged walls follow the equations $z = 0$ ($W_f$ walls), and

$$ex = \pm dz; \; ey = \pm dz \quad (3)$$

($S$ walls), where $e$ and $d$ are the components of the spontaneous strain tensor

$$\begin{pmatrix} -2b & e & e \\ & b & d \\ & & b \end{pmatrix}.$$

Fig. 11 shows an example of permissible configuration of uncharged $S$ walls. Similar to the case of $Pm$, all the other examples can be obtained by rotation about the four-fold axis parallel to [001].

According to Eq. (3) and Fig. 11, in the case of $Cm$ phase all the permissible uncharged walls intersect the (001) plane (the plane of the crystal plate) in the [100] (= $x$) or [010] (= $y$) direction, which is in disagreement with our observations. Therefore we discard the $Cm$ symmetry. On the contrary, in the case of $Pm$ phase, intersections of $S$ walls with the (001) plane are directed along [110] or [1$\bar{1}$0] [see Eq. (1) and Fig. 10]. That is just what we have observed for the directions of the dark stripes between the domains (Figs. 4-6). On the other hand, we have found (see above) that the wide interdomain boundaries that form these stripes can not only be inclined, like the theoretically predicted $S$ walls, but, in some regions of the crystal, also arranged perpendicular to the (001) plane. We will explain this fact later.

The components of the spontaneous strain tensor (2) (i.e. the tensor for the orientation state 1 of Fig. 10) can be calculated using the relations

$$a = (b_m - p)/p, \quad b = (a_m - p)/p,$$
$$c = (c_m - p)/p, \quad 2d = \boldsymbol{p}/2 - \boldsymbol{b}, \quad (4)$$

where $p = (a_m + b_m + c_m)/3$ and $a_m < c_m$. As it comes from the structural diffraction data,[5] this relation between $a_m$ and $c_m$ means that the projection of $\boldsymbol{P}_s$ on $c_m$ axis is larger than on $a_m$ axis [as shown in Fig. 10 (a)].[26] From the values of lattice parameters $a_m = 4.019$ Å, $b_m = 4.006$ Å, $c_m = 4.032$ Å and $\boldsymbol{b} = 90.19^\circ$, measured by x-ray diffraction method for 0.77PMN-0.33PT composition at room temperature,[6] we obtained the following components of the spontaneous strain tensor: $a = -0.0032$, $b = 0$, $c = 0.0032$ and $d = -0.00166$. With these values the domain walls described by Eq. (1) should have an inclination angle $\boldsymbol{q}_1 = 44^\circ$.

As discussed above, the orientation states in a poled crystal could form six types of "elementary" motifs, which are represented in Figs. 10 (b) and (c). Each of these motifs contains two orientation states separated by the walls of the same orientation. In a real crystal all four orientation states are expected, thus the domain structure can be composed of different "elementary" motifs. They can easily be arranged along [001] (vertical) direction so that all domain walls are permissible and uncharged, the examples of which are shown in Fig. 12. But in a horizontal direction (i.e. $\perp$ [001]), different dipole motifs of $S$ walls are incompatible: borders between them should contain nonpermissible or electrically charged walls as shown in Fig. 13 for an example. That is why, when two "elementary" motifs meet, their boundary should be stressed and can deviate from (010) planes to reduce the stresses. Such organization of the boundary can be seen in Figs. 4(b), 5 and 8. The internal stresses or electric charges present at the joint of different motifs induce strain and thereby birefringence which are observed as bright areas while the bulk of the crystal is at the extinction position (see Fig. 8).

Let us now examine the origin of the opaque interdomain stripes and explain why they can have very large and significantly different widths and degrees of darkness (in contrast to usual domain walls) and can be differently inclined. When the 1/2 and 3/4 (or 2/3 and 1/4) motifs alternate along the [001] direction, the inclined $S$ walls are stacked one underneath the other, as shown schematically in Fig. 14. The light is multiply reflected and loses intensity when coming up through the stacks of $S$ walls, so that the corresponding regions appear as comparatively dark stripes directed along <110> when viewed under microscope. The width of these opaque stripes is determined by the angle between (001) plane and the wall, which is the same for all $S$ walls, and by the thickness of vertically alternating domain layers. Thick and thin layers originate wide and narrow opaque stripes, respectively [Figs. 14 (a) and (b)]. Inclined interdomain boundaries (as observed in Fig. 5) appear due to the vertical alternation of the domain layers of the different thickness [Fig. 14 (c)]. It is clear from Figs. 14 (a) and (b) that the wider the interdomain boundary, the smaller the number of the domain walls which are passed by the light. That is why the narrower stripes look darker than the wider ones.

All the alternating domains stacked one underneath the other in the above-described layered structure have mutually perpendicular or the same vibration directions of the slow (and fast) rays, thus they are birefringent objects superimposed in the subtraction or addition positions. The resultant phase difference for the light transmitted through the crystal plate depends on the difference between the net thickness of all domains 1 and 3 and that of all domains 2 and 4 on the path of the light. This difference is a random value and can also be different for the opaque stripes and for the domains between them. Thus the interference color of the crystal observed in the polarizing microscope (which is determined by the phase difference) can vary across the crystal plate as we have indeed observed [see Fig. 4 (a)]. At the same time, the extinction positions of all different domain stacks should be identical and we have really observed simultaneous extinction for the neighboring domains and the opaque stripes between them.

Multiple refraction and reflection on *S* domain walls could cause the rotation of the polarization vector of light transmitting through the crystal. As the density of walls can be different in different areas of the crystal plate, the rotation angle is also different, leading to the observed large-scale spatial variations of the extinction position in the same crystal. However we have observed a clear extinction along [100] (which should be the extinction position for all single domains in the [001]-poled *Pm* crystal plate) in those areas where the opaque interdomain boundaries are vertical (as shown in Fig. 6). This can be explained by the fact that for these regions in between the opaque stripes the light is going through the stacked domains without encountering the *S* walls.

The variation (rotation) of extinction position under an electric drive (Fig. 7) can be explained by the additional birefringence induced via the electro-optic effect. Although in our experiments, because of symmetry arguments, the electric field cannot change the vibration directions for the light traveling along [001] inside a single domain prepoled in the same direction, the optical indicatrix can rotate about the axis lying in the plane of the crystal plate. As a result the conditions of refraction and reflection on the domain walls cange, leading to the additional rotation of the polarization of light and to the change of extinction positions for the multidomain areas.

Note that the orthorhombic *Bmm*2 phase [$m3mFmm2(s)$ species] was discovered in the range of MPB in the other piezoelectric perovskite solid solution, PZN-PT.[5,7] The same orthorhombic phase was also reported for the PMN-PT crystals poled along <011> by a high electric field.[27] This phase can be considered as a particular case of the *Pm* phase under the condition $c_m = a_m$. In the crystal poled along <001> the positions of domain walls do not depend on $c_m$ ($c_m$ parameter is absent in Eq. 1). Consequently, the domain walls directions can be the same in the orthorhombic and the *Pm* phases. The extinction positions in the (001) plate should also be the same. Thus after poling the (001) plate, the orthorhombic phase cannot be distinguished from *Pm* phase by polarizing microscope. Therefore, optical examination of the crystal before poling was an essential step allowing us to conclude that the studied PMN-PT crystal is monoclinic, but not orthorhombic, which is consistent with the published X-ray and neutron diffraction data.

### B. Unpoled crystal

We now calculate the permissible orientations of domain walls in an unpoled monoclinic *Pm* crystal, using the same approach[24] as was applied in the previous section to the poled crystal. All possible variants of the neighboring domain pairs, which can be separated by the permissible $W_f$ and *S* walls and the equations of these walls, are presented in Table II. In this table 84 walls are listed, each of which is characterized by crystallographically different positions and/or different angles between conjugate $P_s$ (and indicatrix axis) directions. Some of these walls have the same directions, so that the total number of crystallographically different positions for the $W_f$ and *S* walls is 45. In addition, 12 walls of the $W_\infty$ type between oppositely polarized domains belonging to the same ferroelastic orientation state can be theoretically expected.

The permissible *S* walls can be divided into three groups depending on the angles they form with the crystallographic directions. We denote the walls belonging to these groups as $S_1$, $S_2$ and $S_3$. The $S_1$ walls defined by *a, b,* and *d* tensor components can be described by Eqs. (1) or equivalent sets of equations listed in Table II. All the *S* walls in the poled crystal are $S_1$ walls. They are inclined at angle $q_1$ to the <100> direction (see Sec. V.A and Fig. 10 for the definition of the angle). But in contrast to the poled crystal, where the charge of the permissible walls is predetermined by their directions, in the unpoled sample any $S_1$ wall can be charged or not, depending on the orientation of $P_s$ in adjacent domains. Head-to-tail orientation provides zero wall charge, while head-to-head and tail-to-tail orientations result in charged walls.

The positions of $S_2$ walls depend on the *a, c,* and *d* values. They are described by the equations

$$(c-a)(x \pm y) = \pm 2dz \qquad (5)$$

or their equivalents (see Table II). Comparing the equations (1) and (5) one can see that they are similar and thus the structural properties of $S_1$ and $S_2$ walls are qualitatively the same. In particular, $S_2$ wall planes are parallel to one of the <110> directions. The difference is quantitative and it lies in the value of the tilting of the walls around these directions. The $S_2$ walls forms the angle of $q_2$, rather than $q_1$, with the corresponding <100> direction.

Equation for $S_3$ walls ($b$, $c$ and $d$ dependent) can be written as

$$x = \pm ny; \quad y = \pm nx, \quad (6)$$

or their equivalents, where

$$n = \frac{2d + [4d^2 + (c-b)^2]^{1/2}}{c-b}. \quad (7)$$

These equations describe the walls between two domains with conjugate $\mathbf{P}_s$ vectors lying within the same {100} plane and making the angle of 90° with each other [e.g. $a$-domains in (001) crystal plate]. These walls are always parallel to the corresponding <100> direction {e.g. $S_3$ walls between $a$-domains are parallel to [001]} and tilted at the angle of $q_3$ to the other <100> direction. This angle approaches 45° when the monoclinic angle $b$ tends to 90° or when the difference between $a_m$ and $c_m$ increases. At $a_m = c_m$ (i.e. in the orthorhombic phase), $q_3$ is equal to zero. In contrast to all other types of walls, $S_3$ walls should not necessarily be neutral even at the "head-to-tail" arrangement of the domains. The neutrality condition is satisfied only when the walls bisect the angle between the conjugate polarization vectors (e.g. vectors 10 and 12), which is in general a quite unlikely event.

Thus, if $S_1$, $S_2$ and $S_3$ walls exist in (001) crystal plate, they can be viewed under the microscope at the angles of $q_1$, $q_2$ and $q_3$ to <100>, respectively. That is why knowledge of these angles is helpful for identifying the domain structure. Using Eqs. (1) and (4) – (6) and the lattice parameters published for the different compositions of the PMN- PT solid solution,[3,5,6] we have calculated the tilting angles. Fig. 15 shows the composition dependences of these angles at different temperatures.

Let us now discuss the origin of domain structure in areas A and B of unpoled crystal (Figs. 2 and 3), where the narrow domains with irregular walls going through the crystal at the angle of $j$ = 24-27° to the [100] direction have been observed. Those domain walls cannot be the crystallographically prominent $W_f$ walls, because all the $W_f$ walls in the $Pm$ phase should be parallel to one of the {100} or {110} planes (see Table II). They cannot be the $S_1$ or $S_2$ walls either, because on the surface of the crystal plate under investigation these walls should look parallel to one of the <110> directions, or they should form an angle with <110> direction (which depends on the lattice parameters). The latter behavior was indeed observed for our sample, but the adjacent domains for $S_1$ or $S_2$ walls should have in this case different extinction positions, as opposite to what we have observed. The same extinction positions should be observed for $a$-domains (90° domains) separated by $S_3$ walls (9/11 or 10/12 walls). Furthermore, from Eq. (6) we calculate the angle $q_3$ = 22° (see also Fig. 15), which is close to the experimental value of $j$ (in Fig. 3). But some other results are incompatible with the existence of these 90° domains separated by $S_3$ walls. Firstly, these walls would most probably be charged. Secondly, examination by a quarter-wave plate (see Sec. III) has indicated that the orientations of indicatrix in all $a$-domains of the same area (A or B) are the same. Therefore, we conclude that the only remaining possibility is the structure of $a$-domains separated by 180° walls inside each area (e.g. 12/12 walls in the area A and 11/11 walls in the area B). This is shown schematically in Fig. 3 (b). Antiparallel domains are not distinguishable in polarized light because the optical indicatrix is invariant under domain (polarization) reversal, but the domain walls between them may be visible.[28] From the point of view of mechanical compatibility, the $W_\infty$ walls can be arbitrary oriented, but to be neutral (in an ideal crystal) they must form cylindrical surface with the generatrix parallel to the $\mathbf{P}_s$ direction. According to the neutron diffraction data[5] obtained from PMNT-PT crystals under the conditions slightly different from our conditions ($x$ = 0.35, $T$ = 80 K) the spontaneous polarization is directed approximately along [103], which corresponds to an angle of 18° with respect to [100], in good agreement with the value of the angle *j* observed in our studies. On the other hand, one can expect that in the vicinity of charged crystal defects the regularity in the orientation of the $W_\infty$ walls may be disturbed. In PMN-PT, differently charged cations ($Mg^{2+}$, $Nb^{5+}$ and $Ti^{4+}$) are known to be randomly distributed on the equivalent crystallographic sites, giving rise to inhomogeneous local electric fields throughout the crystal. We believe that these fields cause the observed microscopic irregularities of the walls in the unpoled monoclinic phase. The presence of the $W_\infty$ walls differently directed with respect to the transmitting light leads to the scattering of light on the walls, rotation of polarization vector and incomplete extinction.

It is worth underlining that the positions of the *S* walls are not noticeably affected by the mentioned above random fields. We have observed plane and regular walls in the poled crystal. This is because their directions are determined by the strain compatibility criteria and the random fields are too weak to disturb them.

As we have already discussed above, the interaction of light with $W_\infty$ walls can lead to the rotatory polarization effects at some angles between walls and incident light vibration direction. The domain motifs of areas A and B are connected by the (010) mirror plane (one of the symmetry elements lost at the cubic-monoclinic transition), thus clockwise polarization rotation in one area (say, A) should be accompanied by the anticlockwise rotation in the other one (B). This consideration is again in full agreement with our observations.

The X-ray diffraction studies on the PMN-PT ceramics revealed the mixture of different phases within the composition range of the MPB.[2,6] In particular, for *x* = 0.33 the coexistence of the monoclinic *Pm* and tetragonal *P4mm* (about 25%) phases was found.[6] Our domain observation and analysis of the unpoled crystals of the same composition clearly indicate the presence of a pure monoclinic phase; the admixture of any other MPB phase (tetragonal, rhombohedral or monoclinic *Cm*) would appear under crossed polarizers as the region of different extinction (if the light propagates through the admixture phase only) or as the region exhibiting no extinction at any positions of polarizers (if the light propagates through the both phases that overlap each other). In PMN-PT crystals, regions without extinction were observed only at the boundaries between areas A and B and can be naturally explained by the overlapping of the monoclinic *a*-domains, i.e. the domains of the same symmetry but oriented at different directions. The shape of A and B areas (and thus the locations of their boundaries) changes significantly after heating/cooling cycles through the Curie temperature, but the same structure of the *a*-domains with distinct extinctions was observed inside A and B areas after several such cycles performed with a virgin or pre-poled crystal, indicating that the phase with the monoclinic symmetry of *Pm* type is indeed a stable single phase. Note that our observations are consistent with the phenomenological model proposed by Cao and Cross[29] who related the multiphase state in the range of MPB to the particle size in a ceramic or powder and predicted that the width of the regions where different phases coexist is inversely proportional to the volume of the particles. Accordingly, in single crystals (which are very large in comparison to the ceramic grains) of the MPB composition, the monophase state is expected to exist.

## VI. CONCLUSIONS

Monoclinic *Pm* phase exhibits stable domain states in $0.67Pb(Mg_{1/3}Nb_{2/3})O_3$-$0.33PbTiO_3$ piezo- and ferroelectric crystals. The domain structure observed appears to be in good agreement with the domain configurations predicted from the analysis of the mechanical compatibility conditions and electrical neutrality of the domain walls for the *m3mFm(p)* species. In the most parts of the unpoled (001) crystal plate we have observed *a*-domains separated by 180° walls (Fig. 3). The positions of the walls are disturbed by the local electric fields resulting from the disorder of the differently charged cations in the crystal structure. As a result, these walls are only approximately parallel to the direction of the spontaneous polarization and look irregular and slightly curved. In addition to these experimentally identified 180° walls, a number of different types of permissible walls are theoretically expected in the *Pm* phase of the unpoled crystal. They are: (i) crystallographically prominent $W_f$ walls parallel to one of the {100} or {110} planes, (ii) $S_1$ walls parallel to one of the <110> directions, (iii) $S_2$ walls that are parallel to <110> as well, but differently inclined, and (iv) $S_3$ walls parallel to <100>. On the surfaces of a {100} oriented crystal, each set of *S*-type walls should appear as straight traces parallel to one of the <110> directions or forming an angle with the <100> direction, which depends on the components of the spontaneous strain tensor. We denote this angle as $q_1$, $q_2$ and $q_3$ for $S_1$, $S_2$ and $S_3$ walls, respectively. The values of $q_1$, $q_2$ and $q_3$ are calculated for PMN-PT as a function of composition at different temperatures (Fig. 15). The $W_f$, $S_1$ and $S_2$ walls are always uncharged if they separate the domains with a "head-to-tail" configuration of spontaneous polarization, but the $S_3$ walls can be uncharged only at some special values of the spontaneous strain tensor. It seems that the complex domain structure of the area C in Fig. 2 (not studied in this work) is composed of the different $W_f$ and/or *S*-type walls.

In the (001) crystal plate poled along the [001] direction all the possible types of mechanically unstressed and electrically uncharged domain walls (see Table I) have indeed been observed. The domain structure includes the $W_f$ walls lying within the plane of the plate and the inclined $S_1$ walls parallel to the [110] or [1$\bar{1}$0] direction. The $S_1$ walls are stacked one underneath the other, forming the opaque stripes of several micrometers wide which are parallel to [110] (if formed by $S_1$ walls parallel to [110]) or [1$\bar{1}$0] direction (if formed by $S_1$ walls parallel to [1$\bar{1}$0]). The boundaries can be vertical or inclined with respect to the plane of the crystal plate (Figs. 4-6, 14). They are separated by domains at distances approximately the same as their width. The polydomain blocks formed by the stacking of $S_1$ walls parallel to different directions ([110] or [1$\bar{1}$0]) are not mechanically or electrically compatible, and therefore the strain fields appear at the joints of the blocks, giving rise to strain-induced birefringence (Fig. 8). These stressed domain configurations may be the origin of the crystal cracking during poling.

Both in the poled and unpoled states we have observed the effect of optical rotatory polarization, which is caused by the interaction of the polarized light with the domain walls. As a result of this effect the extinction positions observed under crossed polarizers can be different from those expected for the single domains and the extinction can also be achieved by decrossing polarizes [Figs. 2 (c) and (d)]. The interaction with the walls which can generally have different concentrations in the different parts of the crystal can lead to the spatial variations of the extinction positions across the sample [Fig. 4 (a)]. More experimental and theoretical investigations are required for understanding the light polarization rotation effect. Upon heating the sequence of phase transformation from the monoclinic *Pm* phase into the tetragonal phase and then into the cubic phase has been observed, which is in agreement with the recently revised phase diagram.

## ACKNOWLEDGMENTS

The authors very grateful to Dr. H. Luo for help in crystal preparation. This work was supported by the U.S. Office of Naval Research (Grant No. N00014-99-1-0738).

TABLE I. Characteristics of the permissible domain walls for poled monoclinic *Pm* and *Cm* ferroelectric phases.

| Walls | *Pm* | | *Cm* | |
|---|---|---|---|---|
| | Permissible wall orientation* | Electric conditions | Permissible wall orientation* | Electric conditions |
| 1/2 | $x = y$ | charged | $y = 0$ | charged |
| 1/2 | $(a-b)(x+y) = 2dz$ | uncharged | $ex = dz$ | uncharged |
| 2/3 | $x = -y$ | charged | $x = 0$ | charged |
| 2/3 | $(a-b)(x-y) = 2dz$ | uncharged | $ey = -dz$ | uncharged |
| 3/4 | $x = y$ | charged | $y = 0$ | charged |
| 3/4 | $(a-b)(x+y) = -2dz$ | uncharged | $ex = -dz$ | uncharged |
| 1/4 | $x = -y$ | charged | $x = 0$ | charged |
| 1/4 | $(a-b)(x-y) = -2dz$ | uncharged | $ey = dz$ | uncharged |
| 1/3 | $y = 0$ | charged | $x = -y$ | charged |
| 1/3 | $z = 0$ | uncharged | $z = 0$ | uncharged |
| 2/4 | $x = 0$ | charged | $x = y$ | charged |
| 2/4 | $z = 0$ | uncharged | $z = 0$ | uncharged |

*$a$, $b$, $d$ and $e$ are the components of the spontaneous strain tensor.

TABLE II. Orientations of the permissible $W_f$ and $S$ domain walls for unpoled monoclinic $Pm$ ferroelectric phase.

| Walls | Permissible wall orientations and types of $S$ walls (in brackets)* | |
|---|---|---|
| 1/2 | $x = y$ | $(a-b)(x+y) = 2dz$ $(S_1)$ |
| 2/3 | $x = -y$ | $(a-b)(x-y) = 2dz$ $(S_1)$ |
| 3/4 | $x = y$ | $(a-b)(x+y) = -2dz$ $(S_1)$ |
| 1/4 | $x = -y$ | $(a-b)(x-y) = -2dz$ $(S_1)$ |
| 1/3 | $y = 0$ | $z = 0$ |
| 2/4 | $x = 0$ | $z = 0$ |
| 7/9 | $x = -z$ | $(a-b)(z-x) = -2dy$ $(S_1)$ |
| 5/9 | $x = z$ | $(a-b)(z+x) = 2dy$ $(S_1)$ |
| 5/10 | $x = -z$ | $(a-b)(z-x) = 2dy$ $(S_1)$ |
| 7/10 | $x = z$ | $(a-b)(z+x) = -2dy$ $(S_1)$ |
| 5/7 | $y = 0$ | $z = 0$ |
| 9/10 | $x = 0$ | $y = 0$ |
| 8/11 | $y = z$ | $(a-b)(y+z) = -2dx$ $(S_1)$ |
| 6/11 | $y = -z$ | $(a-b)(y-z) = -2dx$ $(S_1)$ |
| 6/12 | $y = z$ | $(a-b)(y+z) = 2dx$ $(S_1)$ |
| 8/12 | $y = -z$ | $(a-b)(y-z) = 2dx$ $(S_1)$ |
| 6/8 | $x = 0$ | $z = 0$ |
| 11/12 | $x = 0$ | $y = 0$ |
| 5/6 | $x = y$ | $(c-a)(x+y) = -2dz$ $(S_2)$ |
| 6/7 | $x = -y$ | $(c-a)(x-y) = -2dz$ $(S_2)$ |
| 7/8 | $x = y$ | $(c-a)(x+y) = 2dz$ $(S_2)$ |
| 5/8 | $x = -y$ | $(c-a)(x-y) = 2dz$ $(S_2)$ |
| 3/12 | $x = -z$ | $(c-a)(z-x) = 2dy$ $(S_2)$ |
| 1/12 | $x = z$ | $(c-a)(z+x) = -2dy$ $(S_2)$ |
| 1/11 | $x = -z$ | $(c-a)(z-x) = -2dy$ $(S_2)$ |
| 3/11 | $x = z$ | $(c-a)(z+x) = 2dy$ $(S_2)$ |
| 4/10 | $y = z$ | $(c-a)(y+z) = 2dx$ $(S_2)$ |
| 2/10 | $y = -z$ | $(c-a)(y-z) = 2dx$ $(S_2)$ |
| 2/9 | $y = z$ | $(c-a)(y+z) = -2dx$ $(S_2)$ |
| 4/9 | $y = -z$ | $(c-a)(y-z) = -2dx$ $(S_2)$ |
| 1/7 | $y = nz$ $(S_3)$ | $z = -ny$ $(S_3)$ |
| 3/5 | $y = -nz$ $(S_3)$ | $z = ny$ $(S_3)$ |
| 1/5 | $y = z$ | $y = -z$ |
| 3/7 | $y = z$ | $y = -z$ |
| 2/8 | $x = nz$ $(S_3)$ | $z = -nx$ $(S_3)$ |
| 4/6 | $x = -nz$ $(S_3)$ | $z = nx$ $(S_3)$ |
| 2/6 | $x = z$ | $x = -z$ |
| 4/8 | $x = z$ | $x = -z$ |
| 9/11 | $x = ny$ $(S_3)$ | $y = -nx$ $(S_3)$ |
| 10/12 | $x = -ny$ $(S_3)$ | $y = nx$ $(S_3)$ |
| 9/12 | $x = y$ | $x = -y$ |
| 10/11 | $x = y$ | $x = -y$ |

*$a$, $b$, and $d$ are the components of the spontaneous strain tensor, $n$ is defined by Eq. (7).

**Figures captions**

FIG. 1. Schematic diagram illustrating the experimental setup for *in situ* domain observation of (001) crystal plate in polarizing microscope under electric field and the definition of the angles, **f** and **d**, between the polarizer (P), the analyzer (A) and the crystallographic [100] direction.

FIG. 2. Domain structure of the unpoled (001) PMN-PT crystal platelet under polarizing microscope, showing three different areas, A, B and C, characterized by different extinction positions. The angles between polarizer and [100] direction (**d**), and between polarizer and analyzer (**f**) are indicated (see also Fig. 1 for the definition of **f** and **d** angles).

FIG. 3. Enlarged fragment of the domain structure of the unpoled (001) crystal platelet, showing: (a) fine domain structure of adjacent A and B areas, and (b) schematic sketch of possible orientations of the walls and polarization within the domains. The tone balance of the image was modified using the image processing software to improve visibility of the domain walls. The spontaneous polarization, shown by arrows, lie in the (001) plane (the plane of the sheet). The straight lines show schematically the 180° domain walls, which in reality are slightly irregular and arbitrary curved in the depth of the sample.

FIG. 4. Typical fragments of the domain patterns observed after poling. The photographs were taken at (a) $d = 17°$, $f = 72°$ and (b) $d = 45°$, $f = 90°$

FIG. 5. Domain structures in one and the same area of the (001) poled crystal plate observed without analyzer at three different focus distances (50x objective) imaging the upper crystal surface (top photograph) and two inner layers (middle and bottom photographs) at about 40 μm apart from the surface and from each other. The different domain patterns at various depths indicate that the boundaries of the opaque stripes between domains are not perpendicular to the plane of the crystal plate in this particular area.

FIG. 6. Domain structures in one and the same area of the (001) poled crystal plate observed without analyzer at three different focus distances (50x objective) imaging the upper surface (top photograph), a middle layer and the lower surface of the plate (middle and bottom photographs, respectively). The similar domain patterns at various depths indicate that the boundaries of the opaque stripes between domains are perpendicular to the plane of the crystal plate in this particular area.

FIG. 7. Variation of the extinction position of one of the areas of the poled crystal observed in crossed polarizers ($f = 90°$) as a function of a dc field applied along [001].

FIG. 8. Photograph of the poled crystal platelet in polarizing microscope at extinction position ($d = 16°$, $f = 90°$), revealing the strain-induced birefringence at the joint of two intersecting domain blocks.

FIG. 9. Schematic drawing of the possible ferroelastic orientation states (denoted as 1, 2…, 12) in the monoclinic *Pm* phase (unpoled crystal).

FIG. 10. Schematic drawings of (a): the possible orientation states (denoted as 1, 2, 3 and 4), and (b) and (c): the elementary motifs formed by permissible uncharged domain walls of *S* type (b) and $W_f$ type (c), in the monoclinic *Pm* phase of the crystal poled along the [001] (=z) direction. Thick and thin arrows represent the directions of the spontaneous polarization ($P_s$) and the projections of $P_s$ on the {001} pseudocubic planes, respectively. The walls and the motifs are denoted as 1/2, 2/3, 3/4, 1/4, 1/3 and 2/4 in accordance with the directions of the related orientation states. .

FIG. 11. Schematic drawings of (a): the possible orientation states (denoted as 1, 2, 3 and 4), and (b): the example of the elementary motifs formed by permissible uncharged *S* walls, in the monoclinic *Cm* phase poled along the [001] (=z) direction. Thick and thin arrows represent the directions of the spontaneous polarization ($P_s$) and the projections of $P_s$ on {001} pseudocubic planes, respectively.

FIG. 12. Variants of the arrangements along the [001] (vertical) direction for the different elementary motifs of domain walls in the monoclinic *Pm* phase poled along the same direction. All domain walls are permissible and uncharged.

FIG. 13 Variants of the arrangements in the horizontal directions for the different elementary motifs of domain walls in the [001]-poled monoclinic *Pm* phase. Nonpermissible walls and permissible but charged walls are shown by broken and thick lines, respectively.

FIG. 14. Explanation on the formation of interdomain dark stripes of different widths observed in a poled (001) crystal plate. Domain walls are shown by solid lines. Domains of different orientation states are labeled in accordance with Fig. 10. The areas between broken lines indicated by double-end arrows appear under a microscope as opaque stripes, separating birefringent domains. Examples of wide vertical (a), narrow vertical (b) and inclined (c) opaque boundaries are shown.

FIG. 15. Variations of the domain walls angles $q_1$, $q_2$ and $q_3$ in the monoclinic *Pm* phase as a function of composition at 20 K (circles) and at 300 K (triangles), calculated using the lattice parameters of PMN-PT reported for ceramics by Noheda *et al.*[6] (filled circles and triangles) and for crystals by Singh and Panday[3] (open triangles). Crosses represent the angles calculated using the data by Kiat *et al.*[5] for crystals at 80 K.

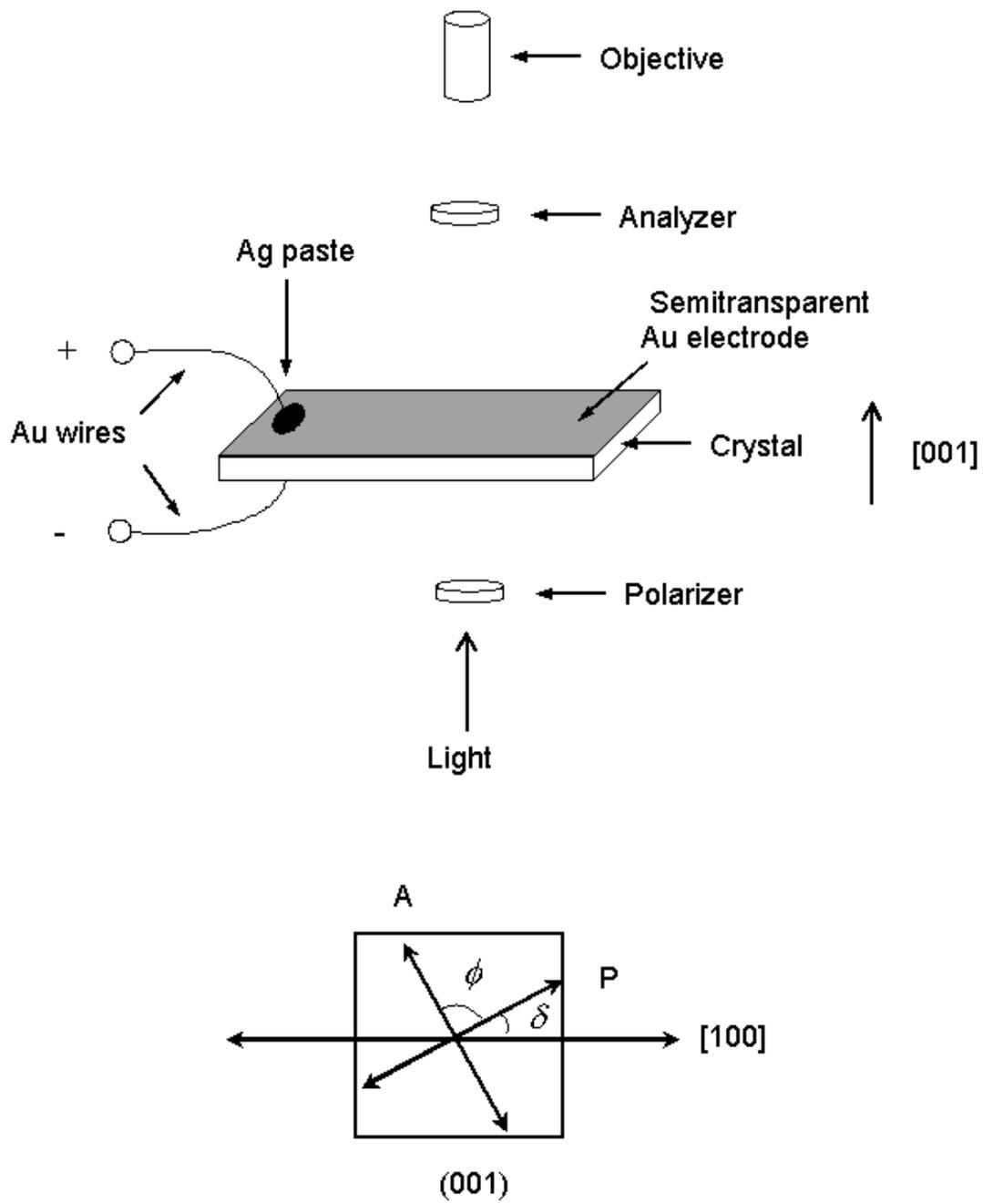

Fig. 1 Bokov & Ye, Phys. Rev. B.

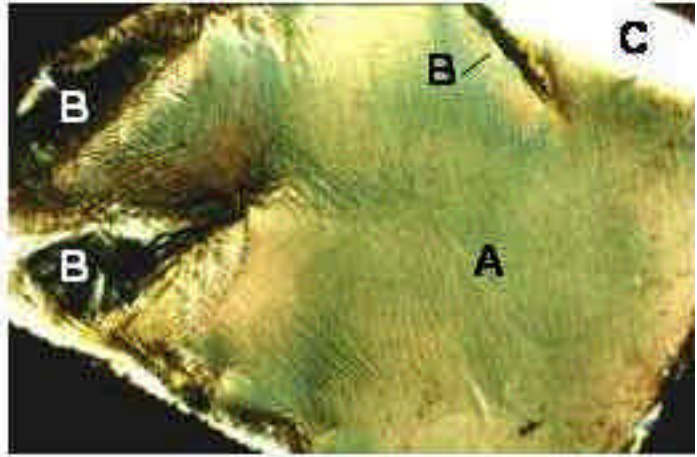

(a)  $\delta = 42°$; $\phi = 90°$

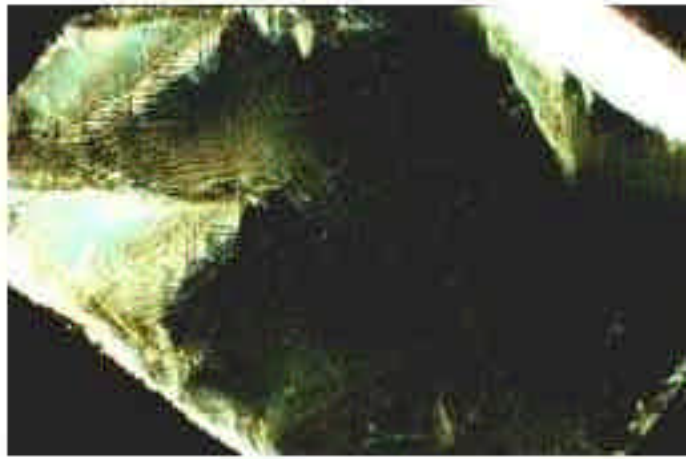

(b)  $\delta = 48°$; $\phi = 90°$

Fig. 2

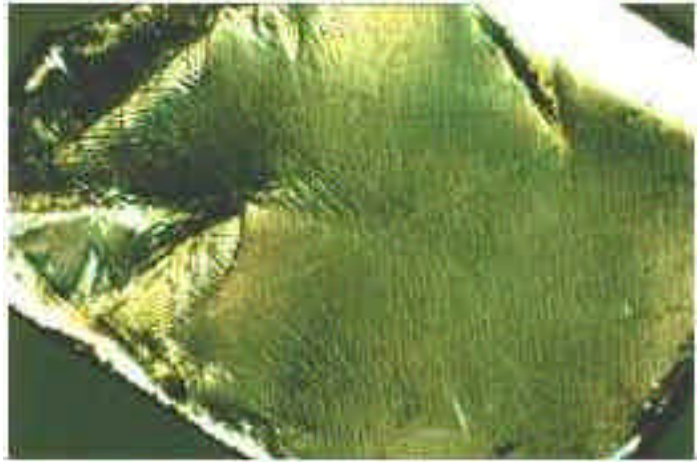

(c)  $\delta = 45°$; $\phi = 87°$

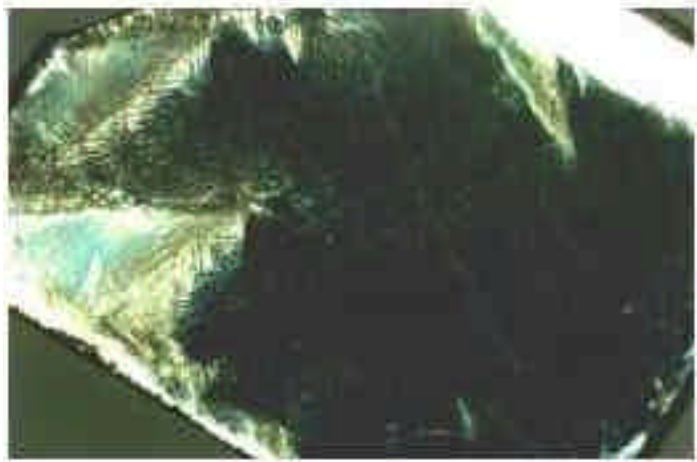

(d)  $\delta = 45°$; $\phi = 93°$

Fig. 2 (Cont)

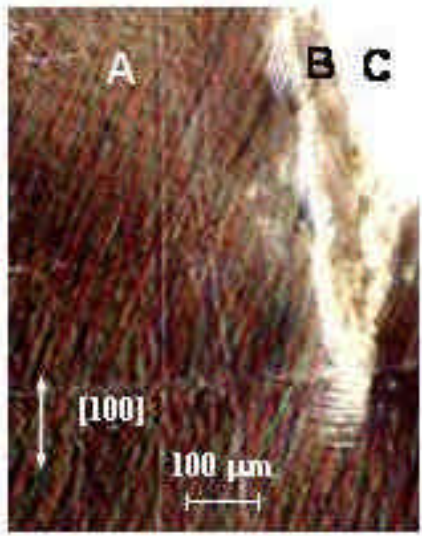 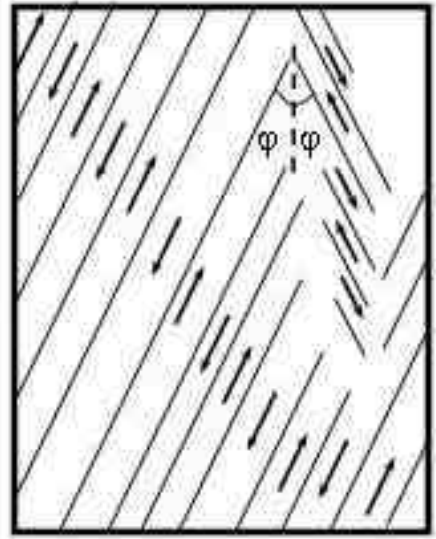

Fig. 3

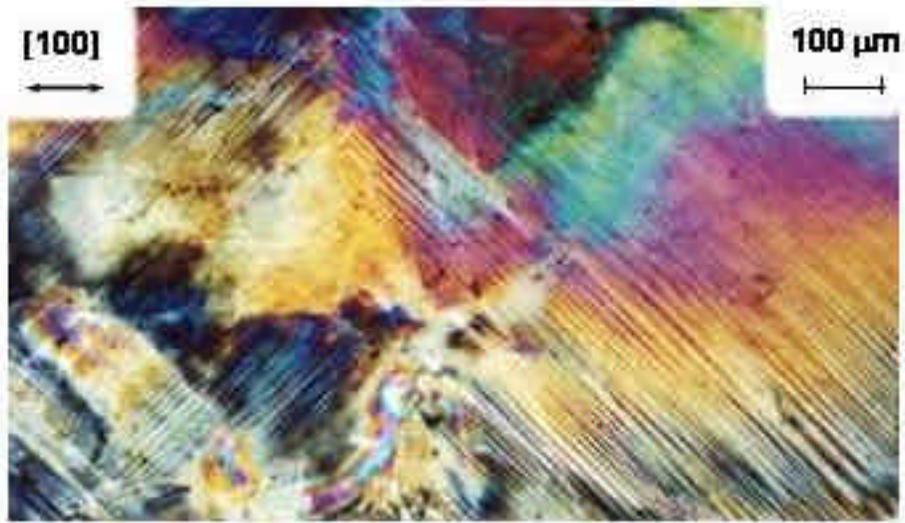

(a)

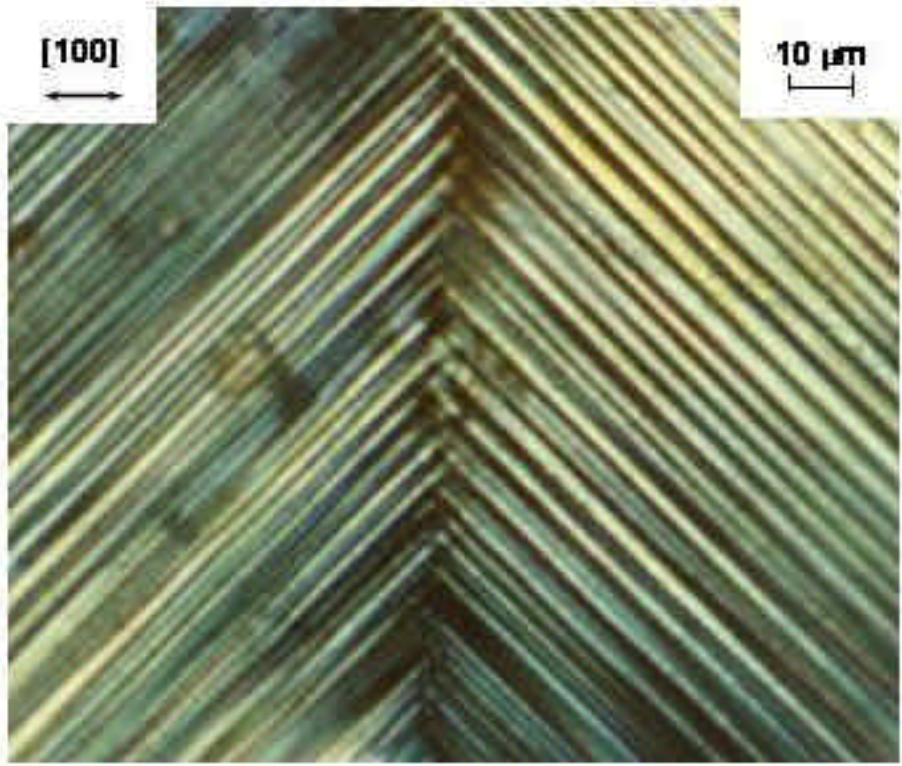

(b)

Fig. 4

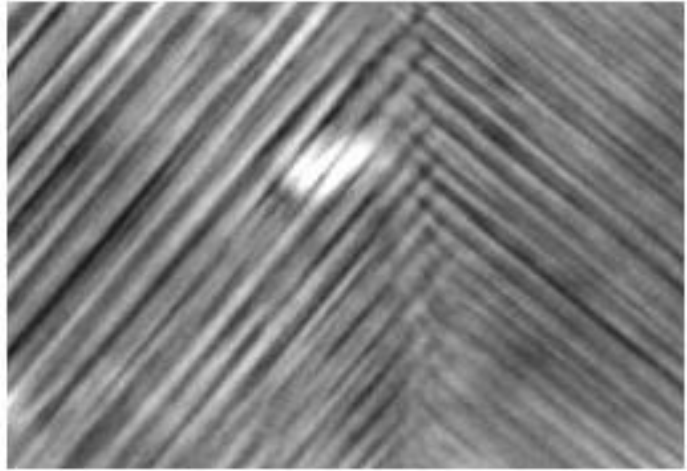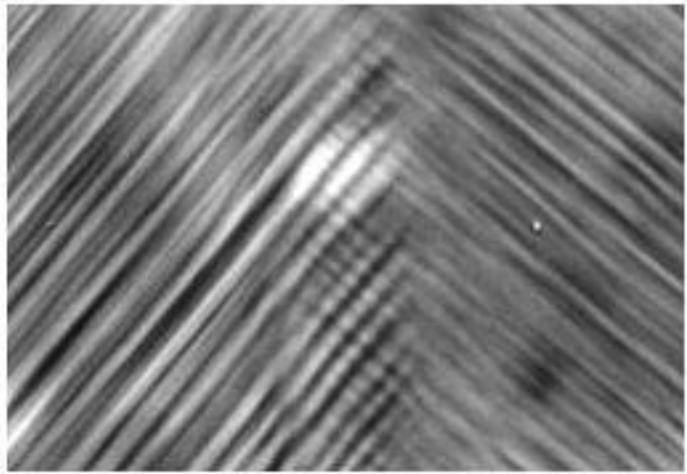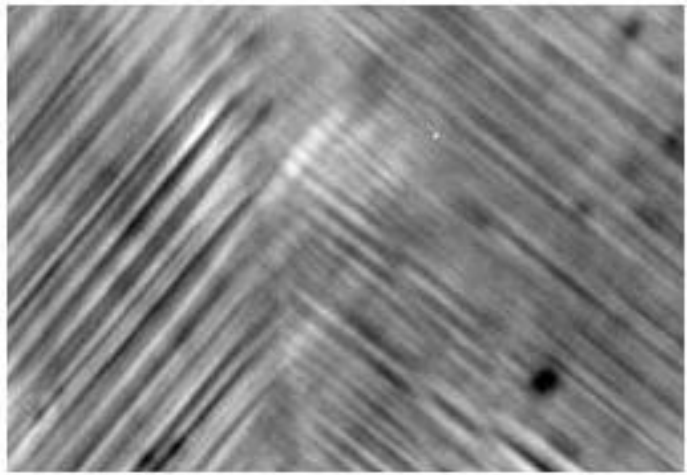

[100]  20 μm

Fig. 5

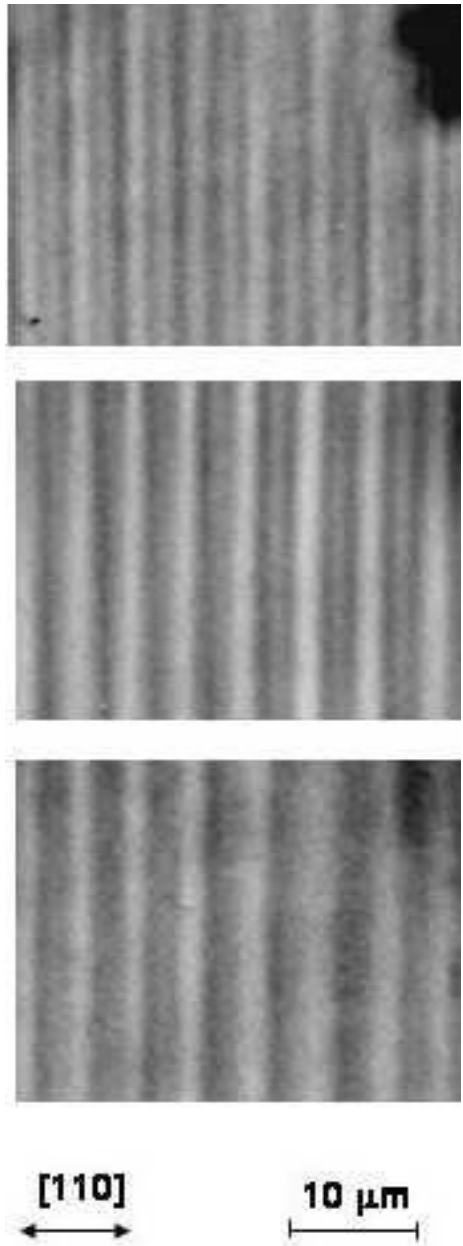

Fig. 6

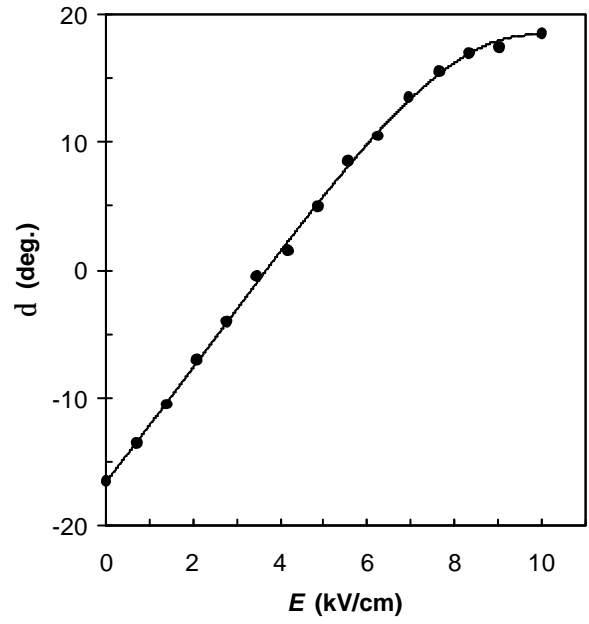

Fig 7 (Bokov & Ye)

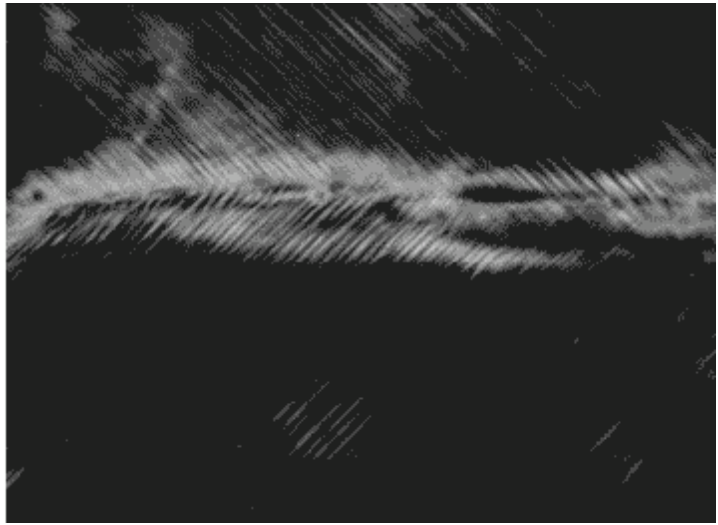

Fig 8 (Bokov & Ye)

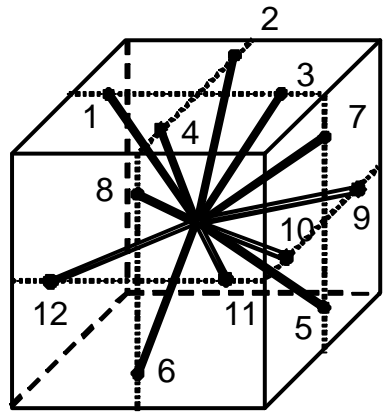

Fig. 9 (Bokov & Ye)

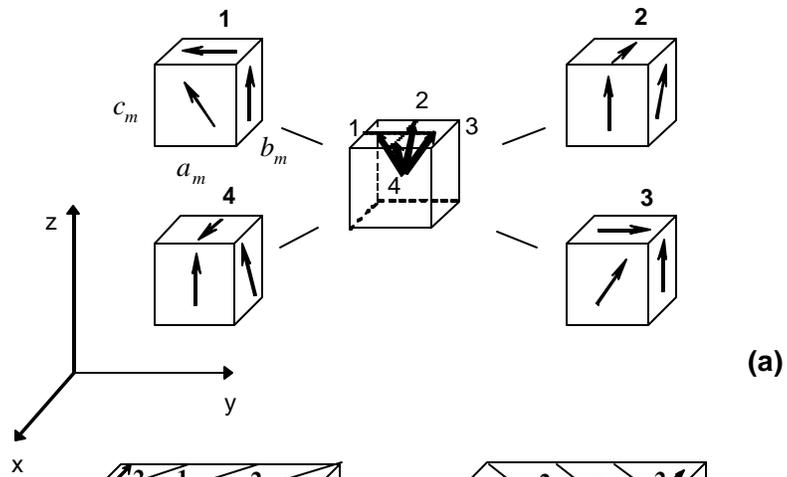

(a)

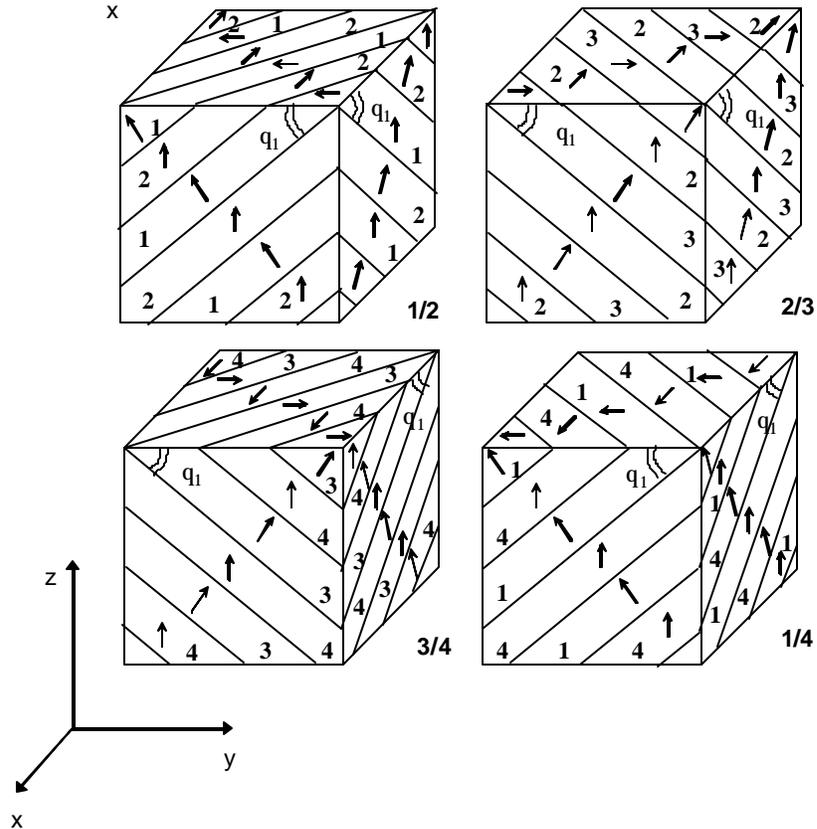

(b)

Fig. 10. Bokov & Ye, *Phys.Rev. B*

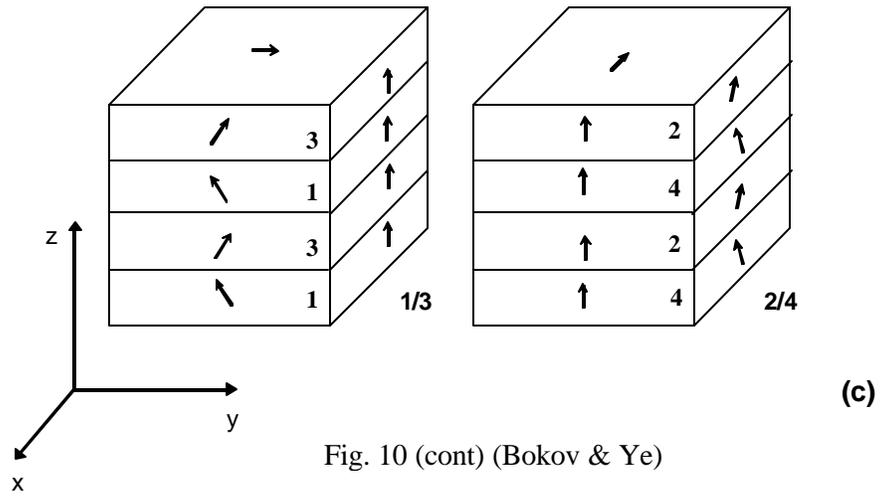

(c)

Fig. 10 (cont) (Bokov & Ye)

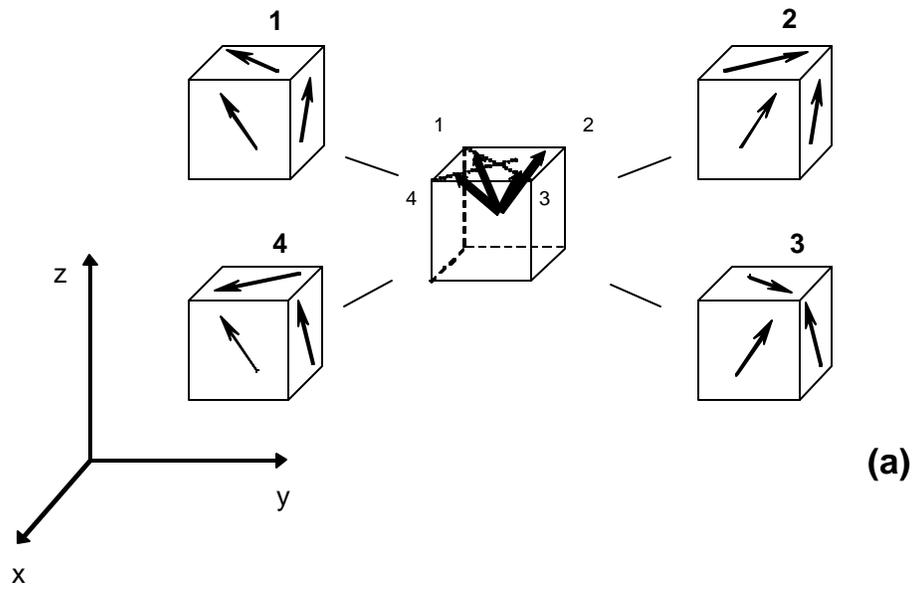

(a)

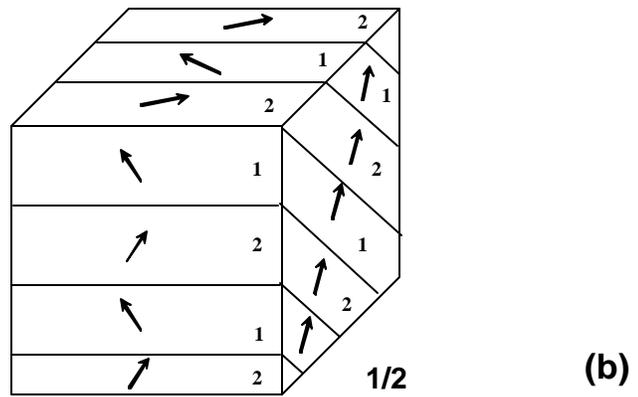

(b)

Fig. 11 (Bokov & Ye)

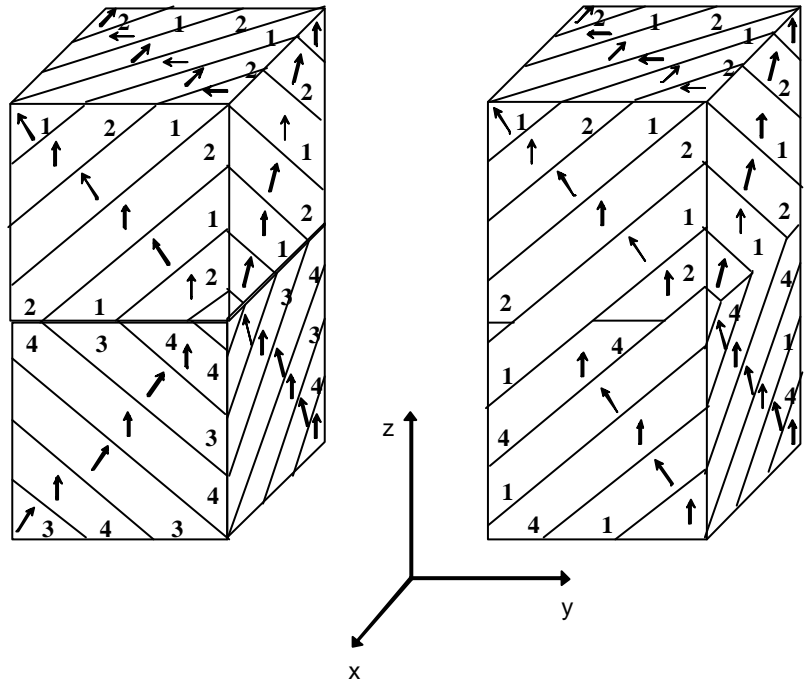

Fig. 12 (Bokov & Ye)

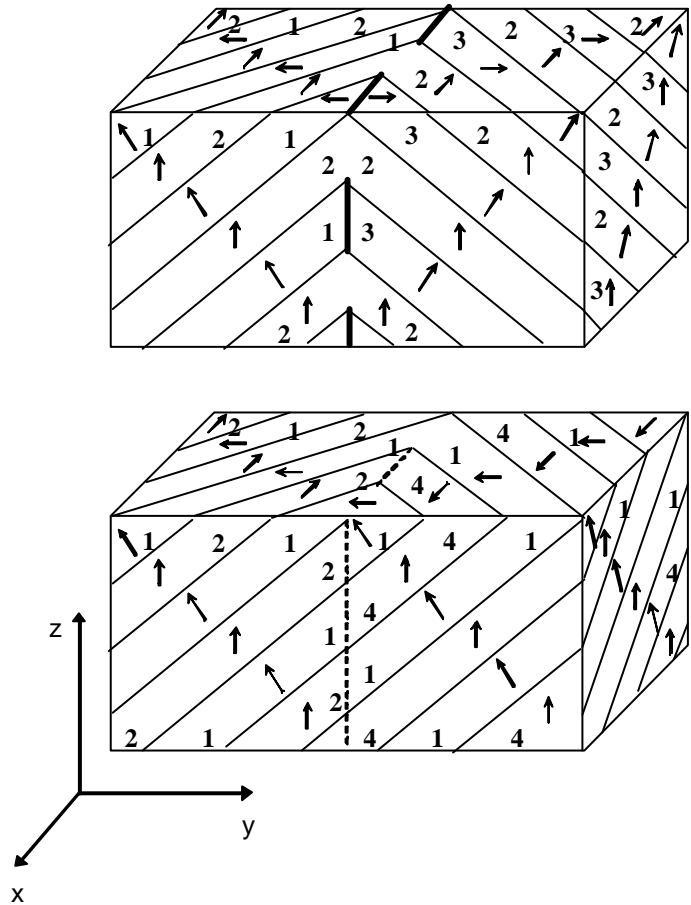

Fig. 13 (Bokov & Ye)

**28**

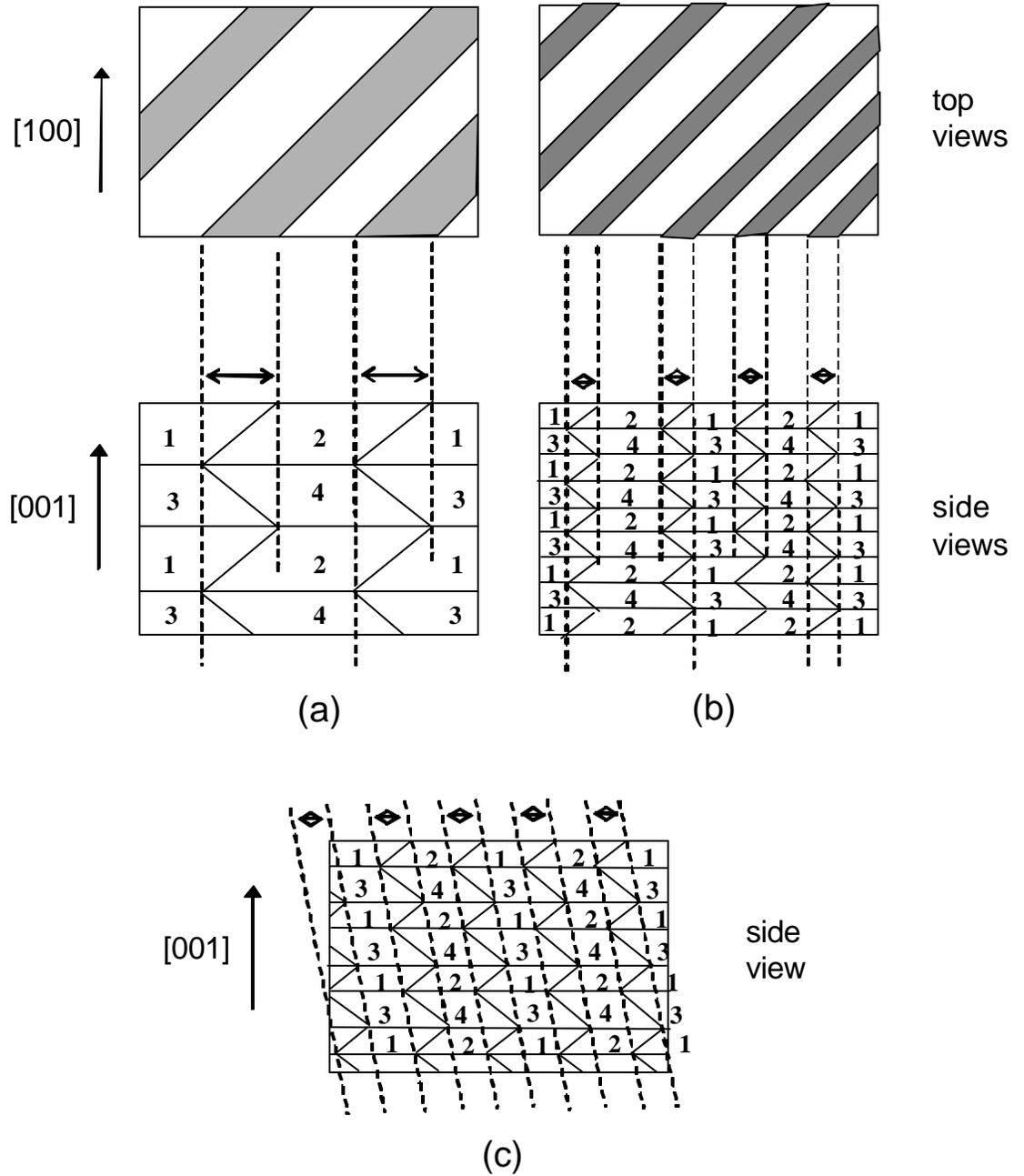

Fig. 14. Bokov & Ye, *Phys. Rev. B*

**29**

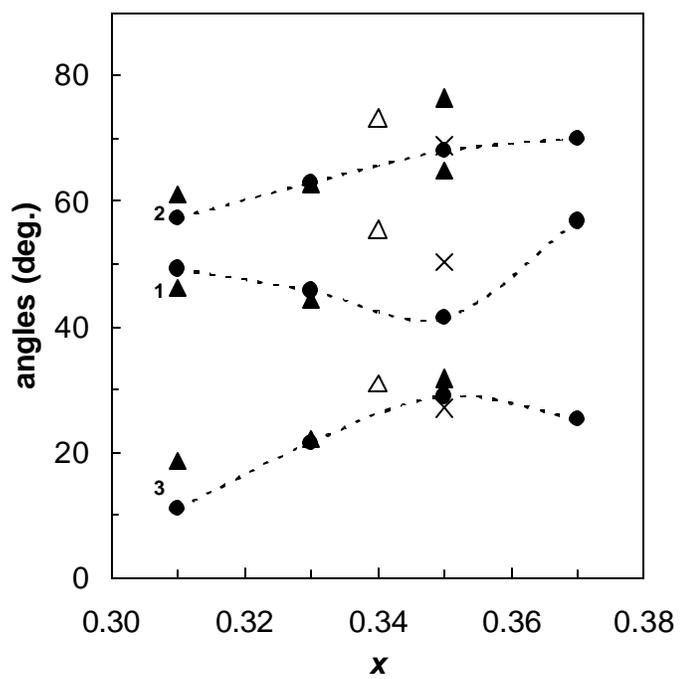